\documentclass[a4paper]{article}

\usepackage[english]{babel}
\usepackage[utf8x]{inputenc}
\usepackage[T1]{fontenc}

\usepackage[a4paper,top=3cm,bottom=3.5cm,left=3cm,right=3cm,marginparwidth=1.75cm]{geometry}

\usepackage{amsmath}
\usepackage{graphicx}
\usepackage{subcaption}
\usepackage{url}
\usepackage[colorlinks=true, allcolors=blue]{hyperref}
\usepackage{enumitem}
\usepackage{xcolor}

\usepackage{array}
\usepackage{tabularx,booktabs}
\newcolumntype{Y}{>{\centering\arraybackslash}X}

\numberwithin{equation}{subsection}

\begin{document}
\title{\textbf{Long Short-Term Memory Neural Network \\for Financial Time Series}}
\author{Carmina Fjellstr\"{o}m}
\date{}

\newcommand{\Addresses}{{
  \footnotesize

  \textsc{Department of Mathematics, Uppsala University, S-751 06 Uppsala, Sweden}\par\nopagebreak
  \textit{E-mail address}: \texttt{carmina.fjellstrom@math.uu.se}

}}

\maketitle

\begin{abstract}
Performance forecasting is an age-old problem in economics and finance. Recently, developments in machine learning and neural networks have given rise to non-linear time series models that provide modern and promising alternatives to traditional methods of analysis. In this paper, we present an ensemble of independent and parallel long short-term memory (LSTM) neural networks for the prediction of stock price movement. LSTMs have been shown to be especially suited for time series data due to their ability to incorporate past information, while neural network ensembles have been found to reduce variability in results and improve generalization. A binary classification problem based on the median of returns is used, and the ensemble’s forecast depends on a threshold value, which is the minimum number of LSTMs required to agree upon the result. The model is applied to the constituents of the smaller, less efficient Stockholm OMX30 instead of other major market indices such as the DJIA and S\&P500 commonly found in literature. With a straightforward trading strategy, comparisons with a randomly chosen portfolio and a portfolio containing all the stocks in the index show that the portfolio resulting from the LSTM ensemble provides better average daily returns and higher cumulative returns over time. Moreover, the LSTM portfolio also exhibits less volatility, leading to higher risk-return ratios. 
\end{abstract}

\section{Introduction}
Prediction of asset prices has long been a central endeavor in mathematical finance and econometrics. Financial time series, however, are notoriously challenging to analyze because of their nonstationarity, nonlinearity, and noise, resulting from the irrational human behavior that drive the data. In the past, methods used are those of traditional nature such as ones based on Autoregressive Integrated Moving Average (ARIMA), Generalized Autoregressive Conditional Heteroskedasticity (GARCH), as well as other stochastic volatility models (see, for example, \cite{Bollerslev, Box, Shumway2017, Tsay2005}). The use of these models often entail making assumptions about the data, its underlying distribution, and the different processes affecting it. Because of these assumptions, these methods often generalize poorly for new, out-of-sample data, even though they fit the current data well and do provide valuable insights into the time series \cite{Tran2018}. Recently, developments in machine learning and neural networks have given rise to non-linear time series models that are increasingly being adapted for financial applications. Support vector machines (SVM), restricted Boltzmann machines (RBM), random forests, gradient boosted trees (GBM), and multilayer perceptrons (MLP) are just some examples of the machine learning models that are being used \cite{Sankar2009, Krauss2017, Patel2015, Takeuchi2013, Luca2016}. Amongst these models, one particular type of machine learning architecture, a recurrent neural network (RNN), has been shown, compared to others, to be better suited for sequential data such as time series. The suitability is due to the feedback loops in RNNs that  allow them to use information not just from the current input, but also from past inputs. This is unlike other neural networks that, in general, process inputs as separate, independent data points. There is however, one major problem with RNNs - their inability to learn long-term dependencies due to the infamous vanishing gradient problem \cite{Bao2017, Bengio1994, Lecun2015}. To address this, the long short-term memory (LSTM) was introduced.

In this paper, an LSTM model is used. A type of RNN, an LSTM also has feedback loops, but moreover, it can also regulate its memory by using a gating mechanism that learns which information to keep, to pass on, and to forget. It is widely used and has been shown to have excellent predictive capabilities in natural language processing, handwriting recognition, image recognition, and image captioning. See, for example, \cite{Byeon2015, Graves2009, Greff2017, Sundermeyer2012, Vinyals2015}. In finance, LSTMs have been increasingly used for time series analysis. For example, applications for price predictions on major stock market indices all over the world such as the S\&P500, Shanghai's SSE Index, India's NIFTY 50, and Brazil's Ibovespa are studied in \cite{Bao2017, Cao2019, Heaton2016, Nelson2017, Selvin2017}. In addition, Tsantekidis et al. \cite{Tsantekidis2017} used LSTMs on Finnish companies to predict price movements through high frequency trading data on a limit order book. Apart from predicting prices, Yeung et al. \cite{Yeung2020} employed LSTMs to detect jumps in the values of different stock market indices, and Xiong et al. \cite{Xiong2015} applied LSTMs on the S\&P500 and Google domestic trends data to forecast price volatility. These are just some examples of LSTM implementations on financial time series showing the neural network to produce promising results. Comparisons with other methods have also been made. Siami-Namini et al. \cite{Siami2018}, for example, compared LSTM with ARIMA for time series forecasting. They not only used data from major exchanges such as the Dow Jones Industrial Average (DJIA) and Nasdaq Composite, but also other economic time series such as the M1 money supply, currency exchange indices, and transportation data. Results from their study show that LSTM forecasts have significantly less root mean square error (RMSE) than those from ARIMA. Fischer and Krauss \cite{Fischer2018} applied LSTM to S\&P500 data for price prediction and compared the results with random forest, a standard deep neural network (DNN), and logistic regression. Their findings indicate that LSTM does indeed have higher accuracy than the other approaches, and that LSTM-based portfolios offer higher returns and lower volatilities. Di Persio and Honchar \cite{Luca2016}, on the other hand, compared LSTM and MLP with their own method, which is an ensemble of wavelets and a convolutional neural network (CNN). Although they reported that their method appears to be superior, the results are very close to those from LSTM \cite{Nelson2017}.

Most of the literature on the application of LSTM on financial time series have been made on major market indices such as the DJIA and S\&P500. In this paper, an LSTM is applied to Stockholm's OMX30 to explore what advantages an LSTM-based approach can provide for a smaller, less perfect market. The method used is inspired by both Fischer and Krauss \cite{Fischer2018} and Barra et al. \cite{Barra2020}. LSTMs are applied to sequences of daily returns; however, as opposed to applying the network to the index itself, the model is applied to the individual stock constituents. Rather than the usual regression problem that is commonly seen in literature, a binary classification problem is used based on the daily median across the different stocks. The target is therefore whether or not the stock's next day return will be above or below the median. Furthermore, instead of just one LSTM, an ensemble of independent and parallel LSTMs is used, where the ensemble's prediction depends on the majority of the individual results. This is in line with \cite{Barra2020} who argue that such ensembles can eliminate much of the randomness in the model and increase the reliability of outcomes. The combination of a median-based binary classification and an LSTM ensemble that this paper implements on the relatively less efficient Swedish market is a unique approach. Results show that, when compared to a randomly chosen portfolio and a portfolio containing all the stocks considered, the LSTM-based method gives rise to portfolios that yield higher returns, lower volatility, and higher risk-return ratios.

The rest of the paper is organized as follows: Section 2 presents LSTMs and their mechanisms in detail. Section 3 describes the method, where the data is introduced and the neural network architecture, ensemble, and trading strategy are explained. Section 4 is the presentation and discussion of results. Finally, section 6 provides a summary and conclusions.

\section{Long Short-Term Memory}
As mentioned above, RNNs are especially suited for sequential data due to their feedback loops that enable them to use both current and past inputs, therefore allowing information to persist. This feature of RNNs means that they are able to learn and take into account trends and context when training and making predictions. There is, however, one major limitation - RNNs lose memory in the long run because of the vanishing gradient problem. To tackle this, Hochreiter and Schmidhuber \cite{Hochreiter1997} introduced the LSTM network in 1997. Since then, it has been modified and improved upon over the years by, for example, \cite{Gers2000, Graves2009, Graves2005,Greff2017}. 

\begin{figure}[hbt!]
  \makebox[\textwidth][c]{\includegraphics[width=1.2\textwidth]{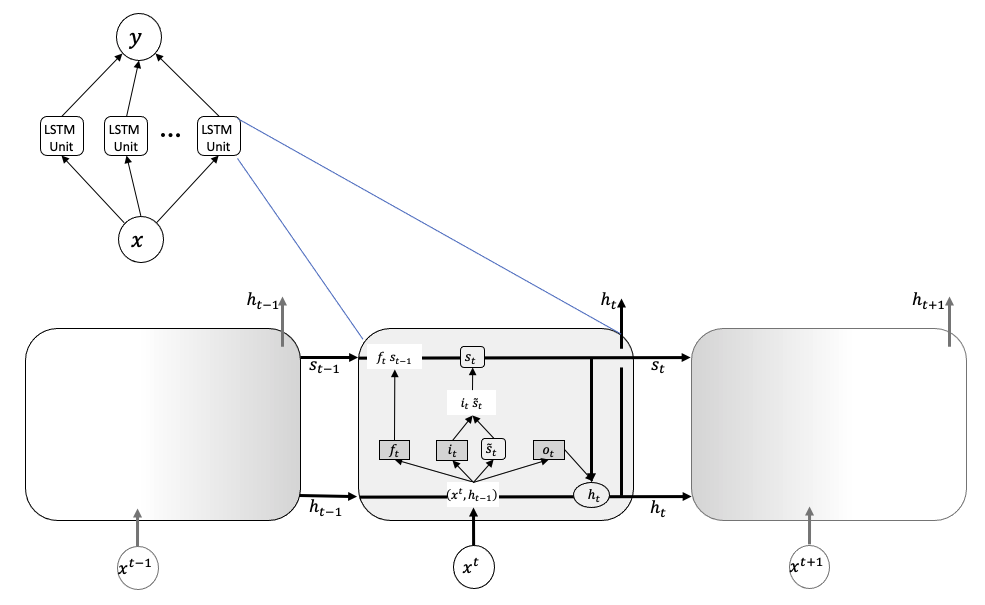}}%
  \caption{LSTM network with a magnified LSTM unit (memory cell)}
  \label{fig:lstm}
\end{figure}

Figure \ref{fig:lstm} shows an example of an LSTM network with one input feature $x$, one hidden layer with several units, and one output $y$. An LSTM unit, also called a \textit{memory cell}, is magnified to show its inner components.
A memory cell contains three gates, each controlling how much information should be kept in memory, forgotten, and passed on as cell output. A $sigmoid$ activation function is used for all three gates as its value ranges from 0, corresponding to no information, to 1, corresponding to all information
\begin{equation}
\sigma(x)=\frac{1}{1+e^{-x}}
\end{equation}

\noindent
The notation in the diagram are as follows:
\begin{itemize}
    \itemsep0em
    \item $x=(x_1,x_2,\ldots,x_n)$ is the input vector where $x_t$, $t=1,\ldots,n$ is the data point at time $t$ in a sequence of length $n$
    \item $s_t$ is the cell state, i.e., the memory of the cell at time $t$
    \item $\Tilde{s}_t$ is the candidate cell state
    \item $h_t$ is the output of the cell, also called hidden state
    \item $f_t$, $i_t$, and $o_t$ are the values for the forget, input, and output gates, respectively
    \item $W_f$, $W_i$, $W_o$, and $W_{\Tilde{s}}$ are the weight matrices associated with the input $x$
    \item $U_f$, $U_i$, $U_o$, and $U_{\Tilde{s}}$ are the weight matrices associated with the output $h_t$
    \item $b_f$, $b_i$, $b_o$, and $b_{\Tilde{s}}$ are the bias vectors
\end{itemize}

When the model is created, the cell states and outputs are initialized, say to $s_0$ and $h_0$, respectively. For a forward pass, an input $x=(x_1,x_2,\ldots,x_n)$ in the form of a sequence is fed into the model, where the memory cells take the data points $x_t$ consecutively to calculate the new cell states and outputs. Figure \ref{fig:lstm}'s magnified memory cell shows its evolution over time as it processes one data point after another: $x_{t-1}$, $x_t$, $x_{t+1}$, \ldots.

The values for the gates at time $t$ are calculated based on the input $x_t$ and the previous output $h_{t-1}$ as follows:
\begin{equation}
f_t=\sigma(W_fx_t + U_fh_{t-1} + b_f)
\end{equation}
\begin{equation}
i_t=\sigma(W_ix_t + U_ih_{t-1} + b_i)
\end{equation}
\begin{equation}
i_t=\sigma(W_ix_t + U_ih_{t-1} + b_i)
\end{equation}

\noindent
Similarly, a candidate cell state $\Tilde{s}_t$ is computed, still based on the input $x_t$ and the previous output $h_{t-1}$, but with a $tanh$ activation function. This represents the new information that the memory cell has received.
\begin{equation}
\Tilde{s}_t=tanh(W_{\Tilde{s}}x_t + U_{\Tilde{s}}h_{t-1} + b_{\Tilde{s}})
\label{eq:candidate_state}
\end{equation}

\noindent
The actual cell state $s_t$ is then calculated based on the previous cell state $s_{t-1}$ and candidate cell state $\Tilde{s}_t$ from (\ref{eq:candidate_state}), where the values of the forget and input gates, $f_t$ and $i_t$, determine how much should be forgotten from $s_{t-1}$  and retained from $\Tilde{s}_t$ :
\begin{equation}
s_t=f_ts_{t-1} + i_t\Tilde{s}_t
\end{equation}

\noindent
Finally, the output $h_t$ is computed based on the value of the output gate $o_t$ and current cell state $s_t$  calculated above:
\begin{equation}
h_t=o_ttanh(s_{t})
\end{equation}

\begin{figure}[hbt!]
\centering
\includegraphics[scale=0.55]{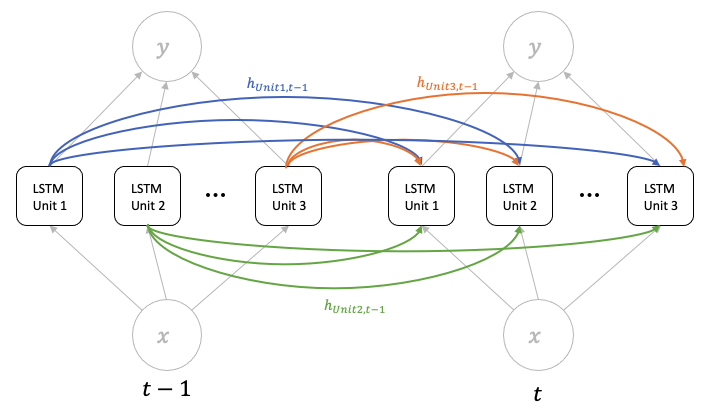}
\caption{\label{fig:hidden}Hidden states $h_t$ from an LSTM unit do not only get looped within the unit, but are also passed on to other LSTM units}
\end{figure}

Note that the outputs $h_t$ are not just looped into the same memory cell, but are also passed to the other memory cells in the network as shown in Figure \ref{fig:hidden}. 

With $m$ being the number of hidden units and $n$ the number of input neurons, the total number of trainable parameters in the neural network is 
\begin{equation}
    4mn + 4m^2 + 4m
\end{equation}
where $4mn$ is the number of weights associated with the input, $4m^2$ is the number of weights associated with the outputs $h_t$, and $4m$ is the number of biases.

\section{Method}
\subsection{Data}
Data for the empirical investigation were taken from the constituents of Stockhom's OMX30 (\url{http://www.nasdaqomxnordic.com}). Table \ref{OMX30Secs} lists the top ten securities in the index by weight, while Table \ref{OMX30Inds} presents the different industries. Both are from December 2019 as that was the latest available information at the time the data were extracted.

\begingroup
\renewcommand{\arraystretch}{1.1}
\begin{table}[hbt]
\centering
\begin{tabular}{llc}
\toprule
Ticker  & Security              & Weight (\%)\\\hline
ATCO A  & ATLAST COPCO          & 7.40\\
HM B    & HENNES \& MAURITZ     & 6.56\\
VOLV B  & VOLVO                 & 6.11\\
ERIC B  & ERICSSON              & 5.91\\
INVE B  & INVESTOR              & 5.49\\
ASSA B  & ASSA ABLOY            & 5.45\\
SAND    & SANDVIK               & 5.40\\
SHB A   & SVENSKA HANDELSBANKEN & 4.56\\
ESSITY B& ESSITY                & 4.54\\
SEB A   & SKAND. ENSKILDA BANKEN& 4.51\\
\bottomrule
\end{tabular}
\caption{OMX30 Top 10 Securities by Weight (as of December 2019)}
\label{OMX30Secs}
\end{table}

\medskip

\begin{table}[hbt]
\centering
\begin{tabular}{lcc}
\toprule
Industry  & Weight (\%)             &  No. of Securities\\\hline
Oil \& Gas          & 0.00    & 0\\
Basic Materials     & 3.27    & 3\\
Industrials         & 37.73   & 10\\
Consumer Goods      & 9.21    & 4\\
Health Care         & 4.07    & 2\\
Consumer Services   & 6.56    & 1\\
Telecommunications  & 6.13    & 2\\
Utilities           & 0.00    & 0\\
Financials          & 22.82   & 6\\
Technology          & 10.21   & 2\\
\bottomrule
\end{tabular}
\caption{OMX30 Industry Breakdown (as of December 2019)}
\label{OMX30Inds}
\end{table}
\endgroup

\begin{figure}[hbt!]
\centering
\includegraphics[scale=0.5]{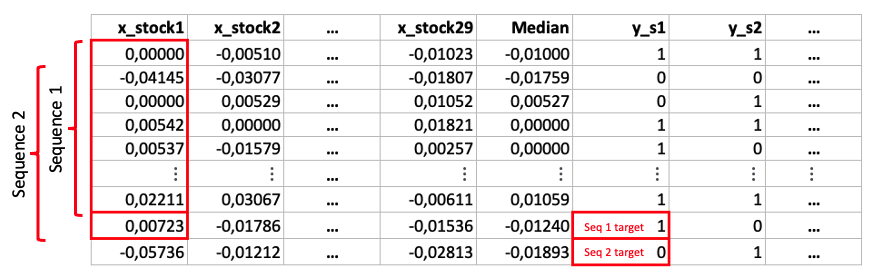}
\caption{\label{fig:sequences}Input sequences and targets}
\end{figure}

Daily closing prices for the constituent stocks\footnote{Essity B excluded due to lack of data for the dates of interest.} from May 2002 to January 2020 were downloaded. In order to avoid keeping track of the changes in constituents over time, the stocks used were kept to be those comprising the index as of February 2020. With the closing prices $p_t$, the daily returns $R_t$ were calculated as follows:
\begin{equation}
    R_t = \frac{p_t}{p_{t-1}} - 1
\end{equation}
In line with \cite{Fischer2018} and \cite{Takeuchi2013}, the daily medians of the returns of the stocks were calculated. Each stock was then classified as either 0 if its daily return is below the daily median, or 1 if its daily return is above. To create the inputs to the LSTM, sequences of returns were created for each stock, where the target for each sequence is the one day ahead prediction of whether the return will be above or below the median. Figure \ref{fig:sequences} illustrates this.

\subsection{Network Architecture}
\label{subsect:network_architecture}
The LSTM model consists of one input neuron, one hidden layer, and one output neuron. A sigmoid activation function is used for the output, which can be interpreted as a measure of confidence. Closer to 1 means that the model is more confident that the return will be above the median, while closer to 0 means it is more confident that it will be below. Adam optimizer was used together with a learning rate of 0.0075, which was chosen with the help of Bayesian optimization \cite{bayes_opt}. The same Bayesian optimization algorithm was also used to determine the values of the other hyperparameters to be:
\begin{itemize}[topsep=0pt]
    \itemsep0em
    \item number of neurons in hidden layer = 3
    \item dropout = 0.06
    \item recurrent dropout = 0.14
    \item batch size = 6800
\end{itemize}

\subsubsection{Training and Testing}
\begin{figure}[hbt!]
  \makebox[\textwidth][c]{\includegraphics[width=1.2\textwidth]{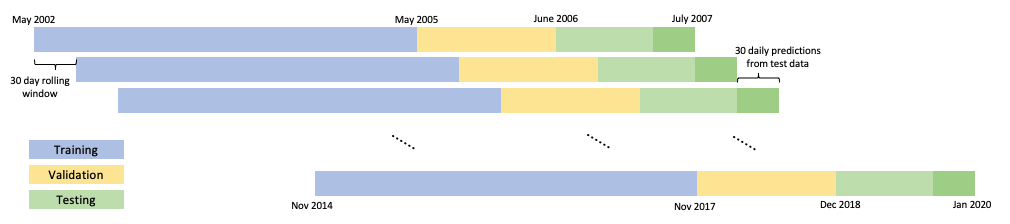}}
  \caption{Blocks of testing, validation, and training sets}
  \label{fig:timeline}
\end{figure}

For training and testing, the data for each stock were divided into blocks of length 750 days for training (approximately three years of trading), 270 days for valuation (more than a year of trading), and 270 days for testing, as illustrated in Figure \ref{fig:timeline}. Sequences of length 240 were created for each of these data sets\footnote{Note that a 270-day testing data with sequences of length 240 will result in 30 days of predictions.}. A rolling window of 30 days was used, which results in 30 non-overlapping days of prediction for each block. In practice, it also means that the model is retrained approximately every six weeks.
\noindent
The 30-day rolling window was chosen based on trial and error, where a shorter rolling window resulted in overfitting, while a longer rolling window resulted in lower accuracy for predictions towards the end of testing as the data becomes further from the training dates.

\subsection{Ensemble and Threshold}
\begin{figure}[hbt!]
  \makebox[\textwidth][c]{\includegraphics[width=1.2\textwidth, height=90pt]{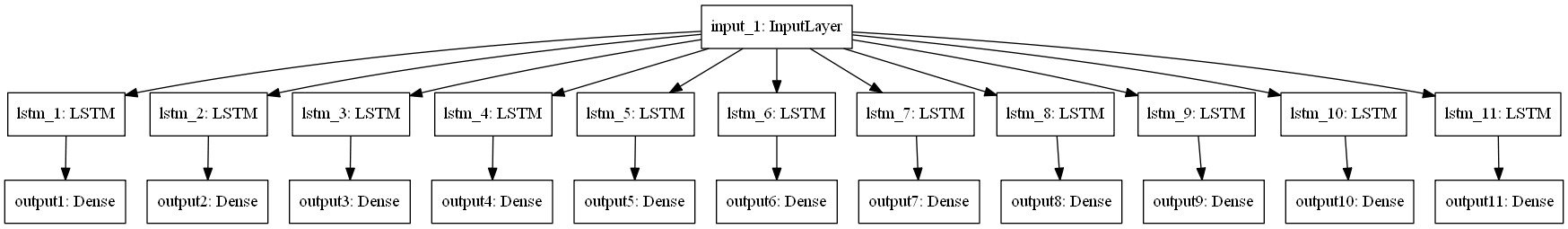}}%
  \caption{LSTM ensemble. The LSTMs are independent of each other and are trained in parallel.}
  \label{fig:model}
\end{figure}

Using ensembles of neural network models has been argued to reduce variation and improve generalization. Therefore, as with \cite{Barra2020}, instead of just one LSTM network, several LSTMs were used, all independent of each other and trained in parallel. A diagram of the ensemble is shown in Figure \ref{fig:model}, where each LSTM has the same architecture described in Section \ref{subsect:network_architecture}, but with a different weight initialization.

\noindent
The different initializations used were those readily available in Keras \cite{keras} and the details of which are provided in Table \ref{initials}. In total, 11 LSTMs were used.

\begingroup
\renewcommand{\arraystretch}{1.2} 
\begin{table}[hbt!]
\centering
\begin{tabular}{p{0.18\linewidth} | p{0.45\linewidth}| l}
\toprule
Initialization  & Description              & Parameters\\\hline
RandomNormal    & Normal Distribution   & $\mu = 0.0$, $\sigma = 0.05$\\
RandomUniform   & Uniform Distribution  & $min = -0.05$, $max = 0.05$\\
TruncatedNormal & Normal Distribution, values $\sigma\ge2$ are redrawn & $\mu = 0.0$, $\sigma = 0.05$ \\
Zeros           & Initialized to 0   \\
Ones            & Initialized to 1 \\
GlorotNormal    & Normal Distributed & $\mu = 0.0$, $\sigma = \sqrt{\frac{2}{fan\_in + fan\_out}}$\\
GlorotUniform   & Uniform Distribution$[-limit,limit]$&  $limit = \sqrt{\frac{6}{fan\_in + fan\_out}}$\\
Identity        & Identity matrix& \\
Orthogonal      & Orthogonal matrix from QR decomposition of random matrix drawn from normal distribution& \\
Constant        & Initialized to a constant& $constant = 0.05$\\
VarianceScaling & TruncatedNormal& $\sigma=\sqrt{\frac{1}{fan\_in}}$\\
\bottomrule
\end{tabular}
\caption{Different weight initializations with chosen parameters for the individual LSTMs} 
\label{initials}
\end{table}
\endgroup

As the LSTMs are independent, so are their outputs. The final ensemble prediction as to whether the next day return will be above or below the median is decided based upon a minimum number, referred to as a threshold, of LSTMs with that result. In other words, the threshold is the minimum number of LSTMs that must agree on a prediction. Since there are 11 LSTMs, any threshold equal to or above six is considered majority. 

\subsection{Trading Strategy}
For the trading strategy, stocks predicted by the LSTM ensemble to perform better than the median are bought and added to an equally weighted portfolio. A stock is held until the model no longer predicts it to be above the median, in which case, the position is closed. The resulting portfolio is rather dynamic and is adjusted daily based on the LSTM forecasts. There is no fixed number of stocks in the portfolio; it contains however many stocks the model predicts to perform well.

\section{Results}
\subsection{Accuracy}
Figure \ref{fig:acc_vs_year} below displays the test accuracy for three of the LSTMs, with Random Normal, Random Uniform, and Glorot Uniform\footnote{Glorot Uniform is the default Keras initialization.} weight initializations. Other LSTMs showed similar results so these were chosen as examples. The accuracies are shown to be just near 50\%, which is consistent with findings in literature (see, for example, \cite{Barra2020, Luca2016, Nelson2017}). Because of the complexity of financial time series, accuracy achieved with machine learning methods is often around 50\%, unlike the much higher accuracies achieved in other areas such as image recognition. 
\begin{figure}[hbt!]
  \makebox[\textwidth][c]{\includegraphics[width=1\textwidth, height=0.5\textheight]{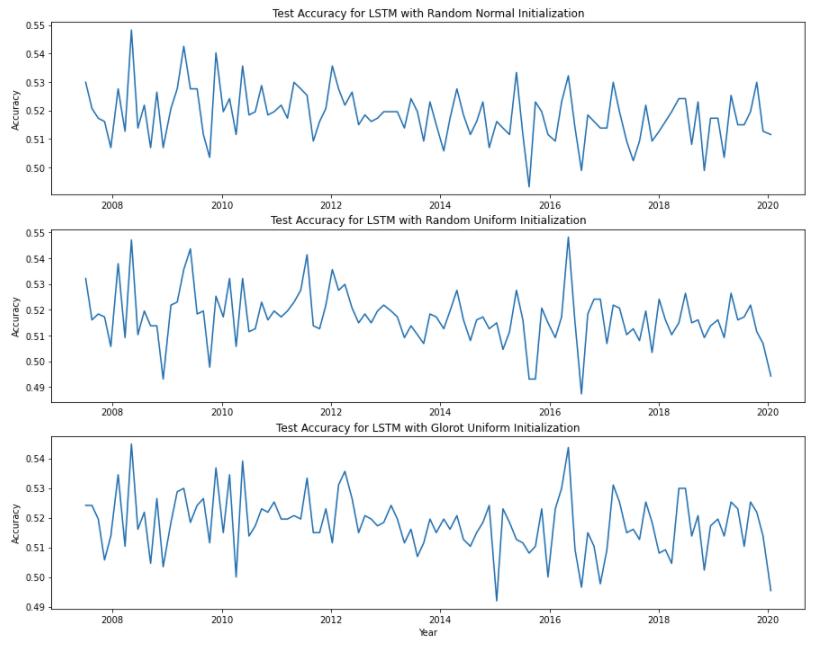}}%
  \caption{Samples of test accuracy for the LSTMS with different weight initializations}
  \label{fig:acc_vs_year}
\end{figure}

\subsection{Threshold and Minimum Accuracy}
\begin{figure}[hbt!]
\begin{subfigure}{.5\textwidth}
  \centering
  \includegraphics[width=1\linewidth]{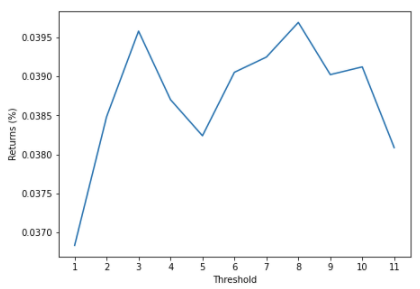}
  \caption{Average Daily Returns (\%) vs. Threshold}
  \label{fig:return_vs_threshold}
\end{subfigure}
\begin{subfigure}{.5\textwidth}
  \centering
  \includegraphics[width=1\linewidth]{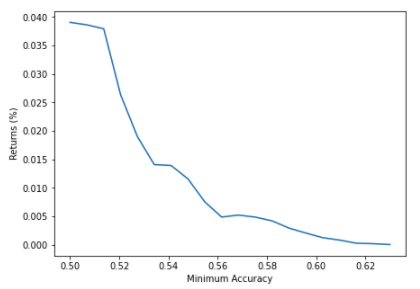}
  \caption{Average Daily Returns (\%) vs. Minimum Accuracy}
  \label{fig:return_vs_acc}
\end{subfigure}
\caption{Average daily returns as threshold and minimum accuracy are varied}
\label{fig:fig}
\end{figure}

To examine the effects of different thresholds and minimum required accuracy on the average daily returns, these variables of the LSTM ensemble were varied. Figure \ref{fig:return_vs_threshold} shows the result for changing the threshold. Excluding three, which appears to be an anomaly, one can see that the daily average portfolio return grows as the threshold increases, until it reaches eight, at which point the return starts to decrease. Having too low of a threshold means that even stocks predicted by many of the LSTMs to perform worse than the median may be added to the portfolio. On the other hand, after the optimal number of eight, requiring more LSTMs to have the same prediction becomes too strong of a condition and results in less or even no stocks included in the portfolio.

\noindent
Similarly, Figure \ref{fig:return_vs_acc} shows the result for varying the minimum required accuracy, where it is shown that the average daily return decreases as the minimum required accuracy increases. An increase in minimum accuracy means requiring the model to be more confident of its prediction. However, as shown above, the predictive accuracy stays very close to 50\%, which means that, as with the threshold, having a higher accuracy requirement may be too strict of a condition, leading to fewer or no stocks being included in the portfolio.

\subsection{Portfolio Returns}
Apart from looking at the effects of the threshold and required accuracy, the daily returns from the LSTM-based portfolio were compared with a portfolio containing all 29 stocks, to be referred to as an all-stock portfolio\footnote{A comparison with the OMX30 index cannot be made directly for two reasons. First, ESSITY B was excluded due to lack of data for the period of interest. Second, the stocks examined here are fixed to be those making up the index as of February 2020, while the index constituents in practice are regularly adjusted.}, and a randomly chosen portfolio. The random portfolio contains randomly chosen stocks from the 29 that were studied and the number of stocks included in the portfolio daily is also random. Table \ref{returns_overall} displays the overall results. 
\begingroup
\renewcommand{\arraystretch}{1.2} 
\begin{table}[hbt!]
\centering
\begin{tabular}{l>{\centering\arraybackslash}p{2.3cm}>{\centering\arraybackslash}p{2.9cm}>{\centering\arraybackslash}p{2.9cm}}
\toprule
		&    LSTM	&	All-Stock Portfolio	&	Random Portfolio\\\midrule
Mean 	&	0,0397	&	0,0336	            &	0,0159	\\
StDev   &	1,2274	&	1,4291	            &	0,7371	\\
Min 	&	-6,1097	&	-8,3910	            &	-4,7493	\\
Max 	&	9,7507	&	9,9950	            &	4,9016	\\
\bottomrule
\end{tabular}
\caption{Comparison of daily returns (\%) for the LSTM, all-stock, and randomly chosen portfolios}
\label{returns_overall}
\end{table}
\endgroup
As seen in the table, both the LSTM and the all-stock portfolios have higher average daily returns than the random portfolio, with the LSTM portfolio having the highest. The LSTM portfolio also has a lower standard deviation than the all-stock portfolio, indicating that the portfolio chosen by the LSTM method may incur less risk. It is also worth noting that although the all-stock portfolio appears to have a slightly higher maximum daily return, the minimum return for the LSTM is less negative, which may suggest that the LSTM portfolio has a much less downside that that of the all-stock portfolio.

For more details, Table \ref{returns_comparison} compares the average daily returns by year, where the best results are highlighted in bold. During the 2007 - 2008 financial crisis, the random portfolio provided better returns than both the LSTM and all-stock portfolios. Afterwards, however, as in the overall results in Table \ref{returns_overall}, the random portfolio is outperformed by the LSTM and all-stock portfolios. The results of these two portfolios are very close to each other, where in some years, the LSTM portfolio appears to have a higher average daily return, while in others, the all-stock portfolio does. Similar to Table \ref{returns_overall}, however, one can see that for periods of negative returns, such as for the years 2007 - 2008, 2011, and 2018, the LSTM appears to do better than the all-stock portfolio by having less negative results. Again, this implies that the LSTM portfolio is certainly a more attractive choice when losses are to be expected.
\begingroup
\renewcommand{\arraystretch}{1.2} 
\begin{table}[hbt!]
\centering
\begin{tabular}{l>{\centering\arraybackslash}p{2.3cm}>{\centering\arraybackslash}p{2.9cm}>{\centering\arraybackslash}p{2.9cm}}
\toprule
Year    & LSTM              & All-Stock Portfolio    & Random Portfolio\\\hline
2007	&	-0,0531	        &	-0,0755	            &	\textbf{-0,0407}	\\
2008	&	-0,0872	        &	-0,1743	            &	\textbf{-0,0641}	\\
2009	&	0,1911	        &	\textbf{0,2322}	    &	0,1184	\\
2010	&	\textbf{0,1052}	&	0,1046	            &	0,0621	\\
2011	&	\textbf{-0,0085}&	-0,0462	            &	-0,0246	\\
2012	&	\textbf{0,0805}	&	0,0667	            &	0,0404	\\
2013	&	0,0639	        &	\textbf{0,0699}	    &	0,0275	\\
2014	&	0,0453	        &	\textbf{0,0519}	    &	0,0253	\\
2015	&	0,0181	        &	\textbf{0,0204}	    &	0,0126	\\
2016	&	0,0142	        &	\textbf{0,0441}	    &	0,0141	\\
2017	&	\textbf{0,0344}	&	0,0344	            &	0,0131	\\
2018	&	\textbf{-0,0403}&	-0,0497	            &	-0,0447	\\  
2019	&	\textbf{0,1095}	&	0,1088	            &	0,0434	\\
\bottomrule
\end{tabular}
\caption{Comparison of average daily returns (\%) by year. Best results are highlighted in bold.}
\label{returns_comparison}
\end{table}
\endgroup

\begin{figure}[hbt!]
\centering
\includegraphics[scale=0.5]{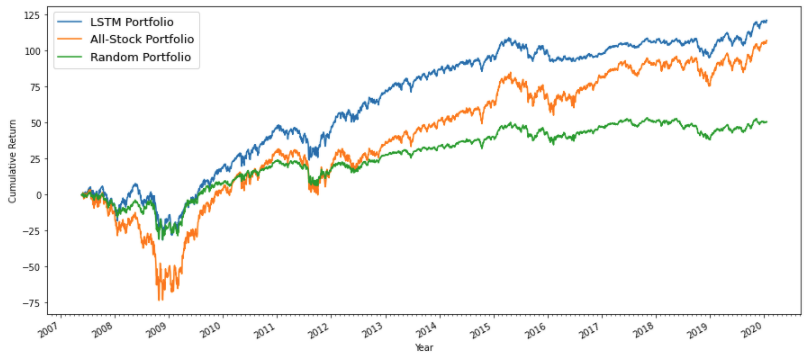}
\caption{Comparison of cumulative returns (\%)}
\label{fig:cum_returns}
\end{figure}

\noindent
To see the development of profits over time, Figure \ref{fig:cum_returns} graphs the cumulative returns for all three portfolios from 2007 - 2020. It can be seen that the LSTM portfolio clearly has the highest accumulated returns for the whole period. Once again, during intervals of decline, as is evident in 2008 - 2009 for example, the graph shows that the LSTM portfolio returns does not decrease as much as the all-stock portfolio.

\subsection{Portfolio Risk}
\noindent
For a more rigorous assessment of the risk, the annualized volatility, Sharpe ratio, and Sortino ratio were examined. The Sharpe ratio is a commonly used measurement of return per unit of risk and is given by the following formula:
\begin{equation}
Sharpe\;ratio=\frac{R_p - r_f}{\sigma_p}
\end{equation}
where the $R_p$ is the portfolio return, $r_f$ the risk-free return, and $\sigma_p$ the standard deviation of the portfolio returns. On the other hand, the Sortino ratio is a variant of the Sharpe ratio that only takes into account the downside standard deviation. It is given by:
\begin{equation}
Sortino\;ratio=\frac{R_p - r_f}{\sigma_d}
\end{equation}
where $\sigma_d$ is the standard deviation of the negative portfolio returns. By only considering the negative returns, the Sortino ratio distinguishes between good and bad volatility. Regardless of which is used, higher values are preferred for both ratios. In the calculations, the risk-free rate was taken to be the Swedish 1-month Treasury bill (accessed from \url{https://www.riksbank.se/en-gb/statistics/search-interest--exchange-rates/}). Results are presented in Table \ref{risk_table}.

\begingroup
\renewcommand{\arraystretch}{1.2} 
\begin{tabularx}{\textwidth}{c *{6}{Y}}
\toprule
 & \multicolumn{2}{c}{Annualized Volatility} 
 & \multicolumn{2}{c}{Annualized Sharpe Ratio} 
 & \multicolumn{2}{c}{Annualized Sortino Ratio}\\
\cmidrule(lr){2-3} \cmidrule(lr){4-5} \cmidrule(l){6-7}
 Year & LSTM & All-Stock Portfolio & LSTM & All-Stock Portfolio & LSTM & All-Stock Portfolio\\
\midrule
2007&   \textbf{20,5268}	&22,8983	&\textbf{-0,7673}	&-0,9333	&\textbf{-1,1907}	&-1,4488\\
2008&	\textbf{35,0996}	&41,8415	&\textbf{-0,7025}	&-1,1130	&\textbf{-1,2706}	&-1,9361\\
2009&	\textbf{23,3151}	&32,2950	&\textbf{2,0473}	&1,7982	    &\textbf{3,5995}	&3,3155\\
2010&	\textbf{18,6689}	&19,8941	&\textbf{1,3992}	&1,3045	    &\textbf{2,4743}	&2,2925\\
2011&	\textbf{26,0842}	&28,1767	&\textbf{-0,1284}	&-0,4556	&\textbf{-0,2094}	&-0,7234\\
2012&	\textbf{18,3997}	&19,2223	&\textbf{1,0462}	&0,8206	    &\textbf{1,7852}	&1,3671\\
2013&	\textbf{12,4070}	&12,6676	&1,2403	            &\textbf{1,3343}	&2,1114	&\textbf{2,2801}\\
2014&	\textbf{13,1716}	&13,7736	&0,8410	            &\textbf{0,9246}	&1,4495 &\textbf{1,5971}\\
2015&	\textbf{18,5197}	&20,5986	&0,2574	            &\textbf{0,2590}	&0,4248	&\textbf{0,4295}\\
2016&	\textbf{14,8670}	&21,0785	&0,2701	            &\textbf{0,5470}	&0,4108	&\textbf{0,8534}\\
2017&	\textbf{9,9064}	&10,7538	&\textbf{0,9232}	&0,8501	    &\textbf{1,5297}	&1,4159\\
2018&	\textbf{13,6962}	&15,5125	&\textbf{-0,7053}	&-0,7752	&\textbf{-1,1330}	&-1,2500\\
2019&	\textbf{13,8528}	&14,5898	&\textbf{2,0087}	&1,8948	    &\textbf{3,3842}	&3,1913\\
\bottomrule
\end{tabularx}
\captionof{table}{Comparisons of risk measures and risk-reward ratios between the LSTM and all-stock portfolios. Better results are highlighted in bold.}
\label{risk_table}
\endgroup

Looking at the annualized volatility columns, it is evident that the LSTM has lower volatility throughout, denoting a less risky portfolio. Comparing the Sharpe and Sortino ratios, one can see that apart from 2013 - 2016, the LSTM portfolio appears to have a higher return per unit of risk than the all-stock portfolio. Negative Sharpe and Sortino ratios in 2007 - 2008, 2011, and 2018 indicate the portfolios' returns during these periods are less than the risk-free rate (see Table \ref{returns_comparison}). Care must always be taken when comparing negative ratios as they can be misleading. For example, for the same return, a larger standard deviation, i.e., higher risk, will result in a less negative ratio, which may lead some to conclude better performance. In this case, however, the less negative ratios of the LSTM portfolio do indeed come from the combination of higher returns and lower volatility. 

\section{Summary and Conclusions}
In this paper, we present an approach for the prediction of stock price movement by using an ensemble of independent and parallel LSTM neural networks. A binary classification problem based on the median of returns is used and the ensemble’s forecast depends on how many of the LSTMs agree on the same output. The model is applied to the constituents of the relatively smaller and less efficient OMX30 index as opposed to the commonly used major stock market indices such as the S\&P500 and DJIA. A straightforward trading strategy is then implemented based on the LSTM forecasts. Compared to a randomly chosen portfolio and a portfolio containing all the stocks in the index, the LSTM-based portfolio appears to have better average daily returns and a higher cumulative return over time. Even more remarkably, the LSTM portfolio exhibits less volatility throughout the whole period than the portfolio containing all the stocks. The combination of this lower volatility with the higher returns results in higher risk-return ratios.

With such encouraging outcomes, we identify several ways to improve the approach even further. For example, using a learning rate decay might increase speed of convergence and accuracy. Having a more systematic way of deciding when the model should be retrained may also lead to better results and prevent overfitting. For example, statistics from the data may be taken regularly, where retraining is done whenever a significant shift in distribution or parameters is observed.  Several input neurons corresponding to the number of stocks instead of a single one for the whole neural network may force the model to process the stocks simultaneously and learn the correlations between them. Finally, instead of being equal, having a trading strategy that weighs the stocks depending on the model’s confidence may also lead to higher returns. These are just some examples on how to increase the performance of a method which already gives promising results.

\nocite{*}
\bibliographystyle{abbrv}
\bibliography{References}

\Addresses
\end{document}